\documentclass[12pt]{article}
\usepackage{epsfig}

\newcommand{\ba}{\begin{eqnarray}}
\newcommand{\ea}{\end{eqnarray}}
\newcommand{\IM}{\mbox{\rm Im}}
\newcommand{\dis}{\displaystyle}

\def\beq{\begin{equation}}
\def\eeq#1{\label{#1}\end{equation}}
\def\eeqn{\end{equation}}

\def\beqa{\begin{eqnarray}}
\def\eeqa#1{\label{#1}\end{eqnarray}}
\def\eeqan{\end{eqnarray}}

\let\bar=\overbar

\def\Dslash{\not{\hbox{\kern-4pt $D$}}}
\def\dslash{\not{\hbox{\kern-2pt $\del$}}}

\def\msb{{\bar{\ssstyle M \kern -1pt S}}}

\def\Title#1{\begin{center} {\Large {\bf #1} } \end{center}}

\begin{document}

\Title{$\vert V_{us}\vert$ from hadronic $\tau$ decays\footnote{Proceedings of CKM 2012, 
the 7th International Workshop
  on the CKM Unitarity Triangle, University of Cincinnati,
  USA, 28 September - 2 October 2012}}

\bigskip\bigskip


\begin{raggedright}  

{\it Elvira G\'amiz \index{G\'amiz, E.}\\
Department of Theoretical Physics\\
University of Granada / CAFPE\\
E-18071 Granada, SPAIN\\
Email: megamiz@ugr.es\\}

\abstract{We review the status and future prospects of the determination of the CKM 
matrix element $\vert V_{us}\vert$ using inclusive strange hadronic $\tau$ decay data. We also 
review the results for $\vert V_{us}\vert$ extracted from experimental measurements of 
some exclusive $\tau$ decay channels such as $\tau\to K \nu$ and 
$\tau\to \pi \nu$.}\bigskip\bigskip
\end{raggedright}

\vspace*{-0.4cm}

\section{Introduction}

The hadronic decays of the $\tau$ lepton serve as an ideal system to study 
low-energy QCD under rather clean conditions~\cite{Pich:2011cj}. The study of these processes has 
allowed the determination of the strong coupling $\alpha_s$ to a level of precision 
only achieved by lattice determinations: $\alpha_s(m_\tau)=0.334\pm0.014$~\cite{Pich:2011bb}. 
The fact that the strong coupling calculated as such a low scale, $m_\tau$ agrees with 
direct measurements of $\alpha_s(M_Z)$ at the $Z$ peak when run to $\mu=M_Z$, 
provides the most precise test of asymptotic freedom~\cite{PichTAU12}
\ba
\alpha_s^\tau(M_Z^2)-\alpha_s^Z(M_Z^2)=0.0014\pm                                     
0.0016_\tau\pm 0.0027_Z\,.
\ea

This precise determination of $\alpha_s$ has been possible thanks to 
the detailed investigation, both experimentally (by ALEPH and OPAL at LEP, CLEO at CESR, 
and the B factories BaBar and Belle) and theoretically~\cite{Rtauth}, 
of the hadronic decay rate of the $\tau$ lepton
\ba\label{eq:Rtaudef}
R_{\tau} \equiv
\frac{\Gamma[\tau^-\to\,{\rm hadrons}(\gamma)]}
{\Gamma[\tau^-\to e^-\overline{\nu_e}\nu_{\tau}(\gamma)]}
=R_{\tau,V}+R_{\tau,A}+R_{\tau,S}\,.
\ea
The experimental separation of the Cabibbo-allowed decays ($R_{\tau,V}+R_{\tau,A}$) 
and Cabibbo-suppressed modes ($R_{\tau,S}$) into strange particles allows us to 
also study SU(3)-breaking effects through the difference
\ba\label{eq:su3diff}
\delta R_\tau \equiv
\frac{R_{\tau,V+A}}{|V_{ud}|^2}-
\frac{R_{\tau,s}}{|V_{us}|^2}\,.
\ea
Flavour independent uncertainties drop out in the difference in Eq.~(\ref{eq:su3diff}), 
which is dominated by the strange quark mass, providing 
an ideal place to determine $m_s$~\cite{mstaudecays}. This determination, however, is 
very sensitive to the value of $\vert V_{us}\vert$ employed. It appears thus natural to 
turn things around and to determine $\vert V_{us}\vert$ with an input for $m_s$ 
using Eq.~(\ref{eq:su3diff})~\cite{Gamiz:2004ar}
\ba\label{eq:vusfromtau}
|V_{us}|^2 = \frac{R_{\tau,S}^{exp}}
{\frac{R_{\tau,V+A}^{exp}}{|V_{ud}|^2} - \delta R_{\tau}^{theor}}\,.
\ea
The main advantage of extracting $\vert V_{us}\vert$ from hadronic $\tau$ decays via (\ref{eq:vusfromtau}) 
is that $\delta R_{\tau}^{theor}$ is around an order of magnitude smaller than the ratio 
$R_{\tau,V+A}^{exp}/|V_{ud}|^2$, and thus the determination of $\vert V_{us}\vert$ is dominated by 
quantities that are measured experimentally. The final precision that can be achieved is then 
an experimental issue. 

\section{Theoretical calculation of $\delta R_\tau$}

The hadronic decay rate of the $\tau$ in Eq.~(\ref{eq:Rtaudef}) is related to the longitudinal (J=0) and 
transversal (J=1) part of vector and axial-vector two-point correlation functions via
\ba\label{eq:Rtaucorr}
R_\tau \;=\; 12\pi\int\limits_0^{M_\tau^2} \frac{ds}{M_\tau^2}\,\biggl(
1-\frac{s}{M_\tau^2}\biggr)^2 \biggl[\,\biggl(1+2\frac{s}{M_\tau^2}\biggr)
\IM\,\Pi^T(s)+\IM\,\Pi^L(s)\,\biggr] \,,
\ea
where the relevant combination of two-point functions is 
\ba\label{eq:PiJ}
\Pi^J(s)\equiv |V_{ud}|^2
\left\{ \Pi^J_{V,ud}(s) + \Pi^J_{A,ud}(s) \right\}+
|V_{us}|^2 \left\{ \Pi^J_{V,us}(s)
+ \Pi^J_{A,us}(s) \right\}\,,
\ea
with the functions $\Pi^J_{ij}(s)$ given by 
$\Pi^{\mu\nu}_{V,ij}(q)\equiv i \int {\rm d}^4 x \,
e^{i q\cdot x}\langle 0 | T \left([V^\mu_{ij}]^\dagger(x) V^\nu_{ij}(0)\right)
| 0 \rangle$ and 
$\Pi^{\mu\nu}_{A,ij}(q)\equiv i \int {\rm d}^4 x \,                                            
e^{i q\cdot x}\langle 0 | T \left([A^\mu_{ij}]^\dagger(x) A^\nu_{ij}(0)\right)
| 0 \rangle$. The vector and axial-vector currents are defined as  
 $V^{\mu}_{ij} \equiv \overline{q}_i \gamma^\mu q_j$ 
and $A^{\mu}_{ij} \equiv \overline{q}_i \gamma^\mu \gamma_5 q_j$.

Using the analytic properties of the correlators, 
we can rewrite Eq.~(\ref{eq:Rtaucorr}) as a contour integral running counterclockwise around the 
circle $\vert s\vert = m_\tau$ in the complex $s-$plane
\ba \label{eq:RtauD}
R_\tau = -i \pi \oint_{|s|=M_\tau^2}
\, \frac{{\rm d} s}{s} \, \left[ 1-\frac{s}{M_\tau^2}\right]^3
\left\{3 \left[1+\frac{s}{M_\tau^2}\right] D^{L+T}(s)
+ 4D^L(s) \right\}\,,
\ea
where the Adler functions $D$ are defined by
\ba
D^{L+T}(s)\equiv - s \frac{\rm d}{{\rm d} s}
[\Pi^{L+T}(s)] \,;\quad\quad\quad
D^{L}(s)\equiv \frac{s}{M_\tau^2} \frac{\rm d}{{\rm d} s}
[s \Pi^{L}(s)] \,.
\ea
The contributions to $R_{\tau,V+A}$ and $R_{\tau,S}$, which 
enter in the definition of the SU(3) breaking difference (\ref{eq:su3diff}), 
are given by the
terms proportional to $\vert V_{ud}\vert^2$ and $\vert V_{us}\vert^2$ 
respectively in the decomposition of the correlation functions in Eq.~(\ref{eq:PiJ}). 

At large enough euclidean $Q^2\equiv-s$, the correlation functions $\Pi^{L+T}(Q^2)$ 
and $\Pi^{L}(Q^2)$ can be computed using operator product expansion (OPE) techniques 
as a series of local gauge-invariant operators of increasing dimension, times appropriate 
inverse powers of $s$. Performing the complex integration in (\ref{eq:RtauD}) we can 
express the SU(3)-breaking difference as 
\ba
\delta R^{}_\tau =
\frac{R_{\tau,V+A}}{|V_{ud}|^2}-
\frac{R_{\tau,S}}{|V_{us}|^2} = N_c \, S_{EW} \, {\dis \sum_{D\geq 2}}
 \left[ \delta^{(D)}_{ud}
-\delta^{(D)}_{us}\right]\,,
\ea
where the symbols $\delta_{ij}^{(D)}$ stand for corrections of dimension D in the OPE 
which contain implicit suppression factors of $1/m_\tau^D$. 

An extensive theoretical analysis of $\delta R_\tau$ was performed in 
Ref.~\cite{13Moscowproc}. The authors included contributions of dimension two 
(proportional to $m_s^2$) and four in the OPE, and estimates of dimension six operators using 
the vacuum saturation approximation. The authors of Ref.~\cite{Gamiz:2002nu} advocated 
for the use of a phenomenological description of 
the longitudinal component of $\delta R^{}_\tau$, which avoids the large 
uncertainty associated with the bad convergence of the corresponding QCD perturbative corrections. 
Following Ref.~\cite{Gamiz:2002nu} but using the prescription in Ref.~\cite{Moscowproc} 
to treat the perturbative QCD correction 
to the $L+T$ $D=2$, namely, averaging over contour improved and fixed order results for 
the asymptotically summed series and taking half of the difference as the uncertainty 
associated with the truncation of the series, one gets~\cite{Moscowproc}
\ba\label{deltaRth}
\delta R_{\tau,th}  = (0.1544\pm 0.0037) 
+ (9.3\pm 3.4)m_s^2
+ (0.0034\pm 0.0028) = 0.239\pm0.030\,,
\ea
where $m_s$ is the strange quark mass in the $\overline{MS}$ scheme and at a scale 
$\mu=2~{\rm GeV}$. The first term contains the phenomenological longitudinal contributions, 
the second term contains the $L+T$ perturbative $D=2$ contribution, while the last term stands 
for the rest of the contributions. Notice that contributions of dimension 4 and higher 
in the OPE are negligible with current uncertainties in the $D=2$ perturbative series. 
The final number in Eq.~(\ref{deltaRth}) is obtained using the average over lattice determinations 
of the strange quark mass, $m_s^{\overline{MS}}(2~{\rm GeV})=93.4\pm1.1$~\cite{LLVav}. The result 
in Eq.~(\ref{deltaRth}) agrees within errors with those obtained by using different prescriptions 
for the perturbative $L+T$ $D=2$ series in Refs.~\cite{proc07,Maltman2010}. 

Using the result in (\ref{deltaRth}), together with the most recent averages of experimental 
results~\cite{HFAG12}, 
$R_{\tau,V+A} = 3.4671\pm0.0084$ and $R_{\tau,S} = 0.1612\pm0.0028$, and 
$\vert V_{ud}\vert=0.97425 \pm 0.0022$~\cite{Vud08}, we get
\ba\label{eq:Vusresult}
\vert V_{us}\vert = 0.2173 \pm 0.0020_{{\rm exp}} \pm 0.0010_{{\rm th}}= 0.2173 \pm 0.0022\,.
\ea

A sizeable fraction of the strange branching fraction is due to the decay 
$\tau\to K\nu_\tau$, which can be predicted theoretically with smaller
errors than the direct experimental measurements~\cite{Gamiz:2007qs}. If the experimental 
measurements for this decay and for $\tau\to K^-\pi^0\nu_\tau$ and 
$\tau\to\pi^-\bar K^0\nu_\tau$ are replaced by theoretical predictions using kaon branching 
fractions~\cite{Passemartau12}, $R_{\tau,S}$ increases about $2.5\%$ and the CKM matrix 
element is shifted up to $\vert V_{us}\vert = 0.2203\pm 0.0025$.

\section{$V_{us}$ from exclusive $\tau$ decays}

Some exclusive $\tau$ decay channels can also be used for the extraction of $\vert V_{us}\vert$, 
given the value of non-perturbative parameters such as $f_K$, $f_K/f_\pi$, or the vector form 
factor at zero momentum transfer $f_+(0)$. Preliminary results for $f_+(0)\vert V_{us}\vert$, 
using the decay rates $\Gamma(\tau\to K \pi \nu)$ and a parametrization of the relevant form 
factors based on dispersion relations can be found in~\cite{Passemartau12}.

One can also consider the branching ratio
\ba\label{eq:tautoKnu}
B(\tau\to K\nu)\, =\, \frac{G_F^2f_K^2\vert V_{us}\vert^2m_\tau^3\tau_\tau}
{16\pi\hbar}\left(1-\frac{m_K^2}{m_\tau^2}\right)S_{EW}\,,
\ea
and the ratio
\ba\label{eq:tautoKnuoverpinu}
\frac{B(\tau\to K\nu)}{B(\tau\to\pi\nu)}\, =\, \frac{f_K^2\vert V_{us}\vert^2
(1-m_K^2/m_\tau^2)^2}{f_\pi^2\vert V_{ud}\vert^2(1-m_\pi^2/m_\tau^2)^2}
\frac{r_{LD}(\tau^-\to K^-\nu_\tau)}{r_{LD}(\tau^-\to \pi^-\nu_\tau)}\,,
\ea
where the $r_{LD}$'s are long-distance EW radiative correction. The 
non-perturbative physics is encoded in the kaon decay constant $f_K$ and 
the ratio of decay constants $f_K/f_\pi$ respectively, which are calculated 
with high precision using lattice techniques. Using the most recent 
experimental averages for $B(\tau\to K\nu)$ and $B(\tau\to\pi\nu)$~\cite{HFAG12}, the values 
of $r_{LD}$ also from~\cite{HFAG12}, the lattice averages for $f_K$ and $f_K/f_\pi$~\cite{LLVav}, 
and $\vert V_{ud}\vert=0.97425\pm0.0022$~\cite{Vud08},  
one gets~\cite{HFAG12}
\ba
\vert V_{us}\vert_{\tau K} = 0.2214\pm0.0022\, ,\quad\quad
\vert V_{us}\vert_{\tau K\pi} = 0.2229\pm0.0021
\ea

\section{Conclusions}

The SU(3) breaking difference between Cabibbo-suppresed and Cabibbo-allowed 
hadronic $\tau$ decay data, $\delta R_{\tau}$, has the potential to give the most precise 
determination of $\vert V_{us}\vert$. The theoretical error in Eq.~(\ref{eq:Vusresult}) 
is already at the same level of precision as more accurate determinations from $K$ 
leptonic and semileptonic decays. The final uncertainty becomes thus an experimental issue 
and will eventually be reduced with the full analysis of the BaBar and Belle 
data~\cite{NugentCKM12}, and even further at future facilities such as Belle II.

\def\Discussion{
\setlength{\parskip}{0.3cm}\setlength{\parindent}{0.0cm}
     \bigskip\bigskip      {\Large {\bf Discussion}} \bigskip}
\def\speaker#1{{\bf #1:}\ }
\def\endDiscussion{}

\end{document}